\documentclass[article,fleqn,usenatbib]{mnras}

\usepackage{graphicx}	
\usepackage{amsmath}	
\usepackage{mathtools}
\usepackage{amssymb}	
\usepackage{natbib}
\usepackage[flushleft]{threeparttable}
\usepackage{hyperref}

\usepackage[T1]{fontenc}
\usepackage{ae,aecompl}

\title[Extended and broad Ly$\alpha$ emission at z$\sim5$]{Extended and broad Ly$\alpha$ emission around 	a BAL quasar at  $z\sim5$}

\author[M. Ginolfi et al.]{
	M. Ginolfi$^{1,2}$\thanks{E-mail: michele.ginolfi@oa-roma.inaf.it}, 
	R. Maiolino$^{3,4}$,
	S. Carniani$^{3,4}$,
	F. Arrigoni Battaia$^{5}$,\newauthor
	~S. Cantalupo$^{6}$,
	R. Schneider$^{1,2}$
	\\
$^{1}$INAF/Osservatorio Astronomico di Roma, Via di Frascati 33, 00040 Monte Porzio Catone, Italy\\
$^{2}$Dipartimento di Fisica, Sapienza Universit\`{a} di Roma, Piazzale Aldo Moro 5, I-00185, Roma, Italy\\
$^{3}$Cavendish Laboratory, University of Cambridge, 19 J.J Thomson Ave., Cambridge CB3 0HE, UK\\
$^{4}$Kavli Institute for Cosmology, University of Cambridge, Madingley Road, Cambridge CB3 0HA, UK\\
$^{5}$European Southern Observatory, Karl-Schwarzschild-Strasse 2, 85748, Garching, Germany \\
$^{6}$Department of Physics, ETH Zurich, Wolfgang-Pauli-Strasse 27, 8093 Zurich, Switzerland \\
}

\date{Accepted XXX. Received YYY; in original form ZZZ}

\pubyear{2018}

\begin{document}
\label{firstpage}
\pagerange{\pageref{firstpage}--\pageref{lastpage}}
\maketitle

\begin{abstract} 
\newline
In this work we report deep MUSE observations of a Broad Absorption Line (BAL) quasar at $z\sim5$, revealing a Ly$\alpha$ nebula with a maximum projected linear size of $\sim 60$ kpc around the quasar (down to our 2-$\sigma$ SB limit per layer of $\rm \sim9\times10^{-19}~erg~s^{-1}~cm^{-2}~ arcsec^{-2}$ for a 1 arcsec$^2$ aperture).
After correcting for the cosmological surface brightness dimming, we find that our nebula, at $z\sim5$, has an intrinsically less extended Ly$\alpha$ emission than nebulae at lower redshift. 
However, such a discrepancy is greatly reduced when referring to comoving distances, which take into account the cosmological growth of dark matter (DM) haloes, suggesting a positive correlation between the size of Ly$\alpha$ nebulae and the sizes of DM haloes/structures around quasars. 
Differently from the typical nebulae around radio-quiet non-BAL quasars, in the inner regions ($\sim 10$ kpc) of the circumgalactic medium (CGM) of our source, the velocity dispersion of the Ly$\alpha$ emission is very high (FWHM>1000 km s$^{-1}$), suggesting that in our case we may be probing outflowing material associated with the quasar.

\end{abstract}




\section{Introduction}

The circumgalactic medium (CGM) is tightly linked to the evolutionary phases of galaxies, both by hosting the gas reservoir necessary to fuel star formation and also being subject to galactic feedback and recycling. 
Thus, observing the nature of the CGM and studying its interplay with the host galaxy, especially at high redshift, helps us to understand how galaxies evolve and how their evolution can influence (and be influenced by) the surrounding environment. 
\newline
A common observational technique used to characterize the cold gas phase of the CGM has been the analysis of its absorption signatures along background sightlines (\citealp{Hennawi2006, Prochaska2009, Hennawi2013, Farina2013, Bouche2013, Bouche2016, Prochaska2017}) which has provided important statistical constraints on several properties of the CGM (\citealp{Adelberger2005, Steidel2010, Prochaska2013}).
However, this method is limited by the sparseness of bright background sources and by its one-dimensional nature, which precludes from capturing the spatial distribution of the CGM.
\newline
An alternative approach is to map the CGM through direct imaging of the Ly$\alpha$ line. 
Theoretical models suggest that three main mechanisms should be able to generate circumgalactic Ly$\alpha$ emission: cooling radiation of gravitationally heated gas (e.g., \citealp{Haiman2000, Yang2006, Dijkstra2009}), ultraviolet (UV) photons produced through shock mechanisms (\citealp{Taniguchi2000,Mori2004}), and recombination radiation following photoionization (often referred as \textit{fluorescence}) powered by UV sources (\citealp{Cantalupo2005, Geach2009, Kollmeier2010}). 
While the fluorescent signal powered by the diffuse metagalactic UV background (\citealp{Hogan1987, Binette1993, Gould1996, Haardt1996}), with an expected surface brightness (SB) of $\rm SB_{Ly\alpha}\sim10^{-20}~erg~s^{-1} cm^{-2} arcsec^{-2}$ (\citealp{Cantalupo2005, Rauch2008}), is still out of reach for current optical instrumentation (but see \citealp{Gallego2017}), Ly$\alpha$ fluorescence is predicted to be boosted up into the detectable regime in the vicinity of bright ionizing sources, as luminous quasars (\citealp{Rees1988, Haiman2001, Alam2002, Cantalupo2005}).
This theoretical prediction has been confirmed by a number of surveys targeting the fluorescent Ly$\alpha$ emission around luminous and radio-quiet quasars, using narrow-band (NB) filters on 8-meter class optical telescopes (e.g., \citealp{Cantalupo2012,  Cantalupo2014, Martin2014,ArrigoniBattaia2016}) and spectroscopic observations (e.g., \citealp{Christensen2006, North2012,  Herenz2015}).
However, these surveys have revealed giant Ly$\alpha$ nebulae, spanning distances from the quasar larger than 100 physical kpc (pkpc), only in less than 10\% of the targets (e.g., \citealp{Cantalupo2014, Hennawi2015}; maximum projected linear sizes $>300$~pkpc), and emission on smaller scales ($R \lesssim50-60$ pkpc) have been detected only in about 50\% of the cases.
This 50\% detection rate is likely due to a combination of limits of the observational techniques, as for instance NB filter losses, spectroscopic slit losses, point spread function (PSF) losses 
and, most importantly, dilution of the signal of the Ly$\alpha$ line into the continuum flux (both background and from the host galaxy) encompassed by the width of the filter (see also \citealp{Borisova2016} for a discussion). 
\newline
Only recently, extraordinary advances have been made using the new Multi Unit Spectroscopic Explorer (MUSE; \citealp{Bacon2010}), a wide integral field spectrograph mounted on the ESO/VLT. 
The ability of MUSE in overcoming the technical limitations of previous observational techniques, combined with its large field of view ($1'\times1'$) have boosted the detection rate up to 100\%, providing ubiquitous detections of Ly$\alpha$ nebulae with maximum projected linear sizes of  $\sim$100 pkpc around bright quasars at $3<z<4$ in only 1 hour observation per source (see \citealp{Borisova2016} and Arrigoni-Battaia et al., in prep.).
\newline
\newline
In this study we exploit MUSE to extend the analysis of Ly$\alpha$ emission surrounding quasars towards higher redshifts and examine the nearly unexplored (within this context) class of Broad Absorption Line (BAL) quasars (another case has been recently reported by \citealp{ArrigoniBattaia2017}, i.e., SDSS J1020+1040 at $z = 3.167$).
In particular, we report deep ($\sim 4$ h) MUSE observations of J1605-0112, a BAL quasar at $z\sim4.92$.
BALs are typically characterized by deep and blueshifted (by thousands km s$^{-1}$) broad absorption troughs associated with UV resonant lines (typically CIV and SiV) tracing very fast outflows (e.g., \citealp{Weymann1991,Maiolino2004,Dunn2010}), and they are thought to mark a specific phase of galaxy evolution in which the quasar wind is particularly powerful. 
\newline
\newline
We present our results in the following order. 
In Section \ref{sec:data_reduction} we describe our target, the basics steps of the data reduction processes, and the analysis performed to search for extended Ly$\alpha$  emission.
In Section \ref{sec:results} we report the results of this work and in Section \ref{sec:discussion} a discussion of their implications.
Conclusions are summarised in Section \ref{sec:conclusions}. 
Throughout the paper, we assume a $\rm \Lambda$CDM cosmology with $\rm \Omega_m=0.3$, $\rm \Omega_{\Lambda}=0.7$ and $h=70 \,\rm km\,s^{-1}$. 
One arcsec at $z\sim4.92$ corresponds to $\rm \sim 6.33$ pkpc.

\section{Observations and data processing}\label{sec:data_reduction}

J1605-0112, a radio-quiet quasar at $z=4.92$, is one of the most extreme Low Ionization Lines-BAL (LoBAL) quasar in the Universe, both in terms of luminosity and in terms of depth and velocity of the absorption features (see Table \ref{tab:properties_quasar}).
Indeed, the CIV $\lambda$1549 \AA~absorption troughs are blueshifted by more than 30.000 km s$^{-1}$, tracing very high velocity outflowing gas, 
and have a balnicity index\footnote{The balnicity index is similar to the Equivalent Width expressed in km s$^{-1}$, as defined in \cite{Weymann1991}.} of 9300 km s$^{-1}$ (see \citealp{Maiolino2004}).
A full 1D spectrum obtained with MUSE and a discussion on the BAL features are reported in the Appendix \ref{sec:MUSE_spectrum}.
\newline
MUSE observations of J1605-0112 have been acquired as part of the ESO programme ID 095.A-0875 (PI: R. Maiolino) between May and August 2015 at the UT4 ESO/VLT.
The collection of data was split into 5 observing blocks (OBs), each of them composed by 4 exposures of 850 s each, with the exposures rotated with respect to each other by 90 degrees. 
The total integration time on source was $\sim 4$ hours.
Observations were carried out under good seeing conditions (FWHM$\sim0.6$ arcsec) 
and airmass $<1.5$.
We achieve a SB limit of $\rm \sim5\times10^{-19}erg~s^{-1} cm^{-2} arcsec^{-2}$ (1 $\sigma$) for a 1 arcsec$^2$ aperture in a single wavelength layer (i.e., $1.25$~\AA).
The SB limit on a pseudo-NB image of $30$~\AA~(a common size for filters used in the past NB surveys), obtained averaging 24 wavelength layers around the location of the Ly$\alpha$, is  $\rm \sim3\times10^{-18}erg~s^{-1} cm^{-2} arcsec^{-2}$ (1-$\sigma$).

\begin{table}
	\caption{Summary of the properties of J1605-0112}
	\label{tab:properties_quasar}
	\begin{threeparttable}
		\begin{tabular}{ccccc} 
			\hline
			RA & 
			DEC &
			redshift& 
			log($\lambda L_\lambda$)\tnote{a}&
			BI(CIV)\tnote{b}\\
			J2000&
			J2000&
			&
			&
			[km s$^{-1}$]
			\\ 
			\hline 
			16:05:01.2 & 
			-01:12:20.6 &
			4.92 &
			46.46&
			9300$\pm$2000\\ 
			\hline
		\end{tabular}
		\begin{tablenotes}
			\item[a] Rest-frame luminosity $\lambda L_\lambda$ at 1450 \AA~ (\citealp{Maiolino2003}).
			\item[b] Balnicity index of the CIV absorption (\citealp{Maiolino2004}).
		\end{tablenotes}
	\end{threeparttable}
\end{table}

\subsection{Data reduction} \label{sec: Data reduction}
Individual exposures have been processed with basic data reduction techniques using the standard \texttt{ESO-MUSE} pipeline v1.6 (\citealp{Weilbacher2012, Weilbacher2014}).
For each of them we performed bias subtraction, flat fielding, twilight and illumination correction, and wavelength calibration, using the \texttt{MUSE\_BIAS}, \texttt{MUSE\_FLAT} and \texttt{MUSE\_WAVECAL} pipeline recipes. 
Following \cite{Borisova2016} we did not perform the sky subtraction using the pipeline receipt but we postponed this task to a later stage, as discussed below.
We then registered the individual exposures using the position of point sources in the field; in this way we ensure an accurate relative astrometry, as shifts of a few tenths of arcsecs can occur owing to the spatial shifts introduced by the derotator wobble between exposures (\citealp{Bacon2015}).
The next steps of the data-reduction process were performed with \texttt{CubeFix} and \texttt{CubeSharp}, custom tools for improved flat fielding correction and sky subtraction respectively, both parts of the \texttt{CubExtractor} software package (Cantalupo, in prep.).
These tools, extensively described in \cite{Borisova2016} and \cite{Fumagalli2017}, have been specifically developed to improve data quality for the detection of faint and diffuse emission in MUSE datacubes.
As last step, we combined all the corrected and sky-subtracted datacubes of individual exposures using an average 3$\sigma$-clipping algorithm. 
At this point we created a white-light image (obtained by collapsing the datacube along the wavelength direction) of the combined datacube in order to identify the continuum sources.
We used the positions and spectra of the detected continuum sources to perform another iteration of flat fielding correction and sky subtraction on individual exposures, before combining them again.
One further iteration is enough to improve substantially the removal of self-calibration effects and data quality.
\newline
The final datacube samples the instrument field of view (FOV) of 1x1 arcmin$^2$ in pixels of size 0.2 arcsec and, for each pixel, contains a spectrum covering the wavelength range $4750-9350$\AA~ in spectral bins of $1.25$\AA~. \newline
In the Appendix \ref{sec:white-light} we show the whole 1x1 arcmin$^2$ field of view of the final datacube. 
To check the accuracy of flux calibration we compared our final product with photometric data from the Sloan Digital Sky Survey (SDSS; \citealp{Eisenstein2011}) without finding any flux deviation within the errors for the few sources in common.
\newline
\newline
%
\subsection{Data Analysis}\label{sec:data_analysis}

\subsubsection{PSF and continuum subtraction}\label{sec:psf_cont}
Revealing extended Ly$\alpha$ structures around a quasar requires a proper subtraction of the PSF.
Generally, accurate PSF removal is a substantial issue in the case of NB imaging  and long-slit spectroscopy  technique (e.g., \citealp{Moller2000a, Moller2000b, Weidinger2005}), but here we take advantage of the Integral Field Spectroscopy (IFS), which provides us images of the quasar and its surrounding at different wavelengths, including spectral regions where no extended line emission is expected, thus allowing for a careful PSF modelling.
\newline
To subtract the PSF contribution we adopt a purely empirical method successfully tested in \cite{Borisova2016}, using \texttt{CubePSFSub}, part of the \texttt{CubExtractor} package (Cantalupo, in prep.).
The algorithm consists in producing a pseudo-NB image with a spectral width of 150 spectral pixels ($\sim 187$\AA) for each wavelength layer.
In this way we construct empirical PSF images to be rescaled and subtracted at each wavelength layer.
We rescale the flux in each empirical PSF image assuming that the quasar dominates the flux within a central region of 5x5 pixels (corresponding to 1x1 arcsec$^2$).
The rescaling  factor between the flux in each layer and the empirical PSF images is computed using an averaged-sigma-clip algorithm in the central 5x5 pixels region, to avoid any possible contamination by cosmic rays or other sources of noise in the individual pixels.
We then cut from the rescaled empirical PSF images a central circular region with a radius of  $\sim5$ times the seeing. 
As a last steps, we mask any pixel with negative flux from this circular cut-out region, and we subtract it from the corresponding  layer in the datacube.
The spectral width of $\sim 187$\AA~is a good compromise between minimizing the PSF variations within the waveband and maximizing the signal-to-noise ratio (SNR) in the empirical PSF images. 
We iterate the PSF removal procedure, masking the spectral interval associated with the nebulae to avoid over-subtraction.
We also remove and mask continuum sources around the quasar, to be not affected by their residuals. 
\newline
We note that, by construction, our PSF subtraction has uncertainties in the 1x1 arcsec$^2$ region used for the PSF rescaling, in particular due to the poor pixel sampling in the central region, where the nuclear emission has a very steep gradient. 
For this reason, any data on such scales are considered not reliable in the remaining of this work.
We also note that our determination of the PSF may be potentially contaminated by the emission from the host galaxy in the central region used for the PSF modelling (see e.g., \citealp{Fynbo1999,Fynbo2000} and \citealp{Zafar2011} for a discussion). 
Indeed, the continuum emission from the host galaxy may affect the rescaling factor and cause systematic effects in the PSF subtraction.
However, in our case, the high luminosity of the quasar ($\lambda L_{1450 \rm{A}} \sim 3\times10^{46}~\rm{erg}~\rm{s}^{-1}$; AB magnitude in the \textit{i}-band, \textit{i}-mag = 19.75
\footnote{Computed from our MUSE datacube with an aperture of 3 arcsec in diameter, assuming a SDSS i-band filter. This value is not corrected for intrinsic absorption.}%
) in comparison to the expected luminosity of its host-galaxy at $z\sim5$, motivate our assumption that the quasar dominates the flux budget within the 1x1 arcsec$^2$ used for the PSF rescaling.
In addition, as discussed in the Appendix \ref{sec:MUSE_spectrum}, within the deep CIV trough, we do not see any evidence for a prominent emission from the host galaxy, further supporting that the host galaxy does not contribute significantly to the observed emission.
\newline
As shown in the next Section, our empirical PSF subtraction method produces good results, unveiling low-SB flux around the quasar on large scales.
\newline
After quasar PSF subtraction has been performed, we remove any other possible continuum sources for each spaxel in the cube using a fast median-filtering approach, following \cite{Borisova2016}.

\subsubsection{3D detection and extraction}
We use the PSF-subtracted and continuum-subtracted cube for the detection and the  extraction of extended line emissions.
This task has been performed using  \texttt{CubExtractor} (Cantalupo, in prep.), a 3D automatic extraction software based on a 3D extension of the connected-labelling-component algorithm with union finding of classical binary image analysis (e.g. \citealp{Shapiro2001}). 
\newline
We first extracted a sub-cube ($\Delta \lambda \sim 1000$~\AA) from the processed datacube with wavelength range containing the expected Ly$\alpha$ line and, before searching for detection, we apply a spatial gaussian filtering of 0.5 arcsec (without smoothing in wavelength) to bring out extended but narrow features.
An object is detected if it contains a number of connected voxels (individual spatial and spectral elements in MUSE cube) above a user-defined threshold (we used a threshold of 10.000 connected voxels) above a minimum SNR of 2.5, after datacube being smoothed.
The 3D-segmentation-mask produced by \texttt{CubExtractor} is then used for photometry, optimally-extracted images, velocity and velocity dispersion maps.

\section{Results}\label{sec:results}

\begin{table}
	\caption{Properties of the Ly$\alpha$ nebula around J1605-0112}
	\label{tab:properties_nebula}
	\begin{threeparttable}
		\begin{tabular}{ccccc} 
			\hline
			Size\tnote{a}& 
			$\Delta \lambda$\tnote{b} &
			Luminosity & 
			FWHM\tnote{c} &
			CIV / Ly$\alpha$\tnote{d}
			\\
			(kpc)&
			(\AA)&
			(erg s$^{-1}$)&
			(km s$^{-1}$)&
			(2$\sigma$)
			\\ 
			\hline 
			50-60& 
			38.75&
			4.4$\times$10$^{43}$&
			900&
			<0.06\\ 
			\hline
		\end{tabular}
		\begin{tablenotes}
			\item[a] Maximum linear projected size obtained from the 3D-mask.
			\item[b] Maximum spectral width obtained from the 3D-mask, and used for the pseudo-NB image.
			\item[c] Spatially averaged FWHM.
			\item[d] Limit on the CIV/Ly$\alpha$ line ratio (2$\sigma$).
		\end{tablenotes}
	\end{threeparttable}
\end{table}

\begin{figure}
	\centering
	\includegraphics[width=1\columnwidth]{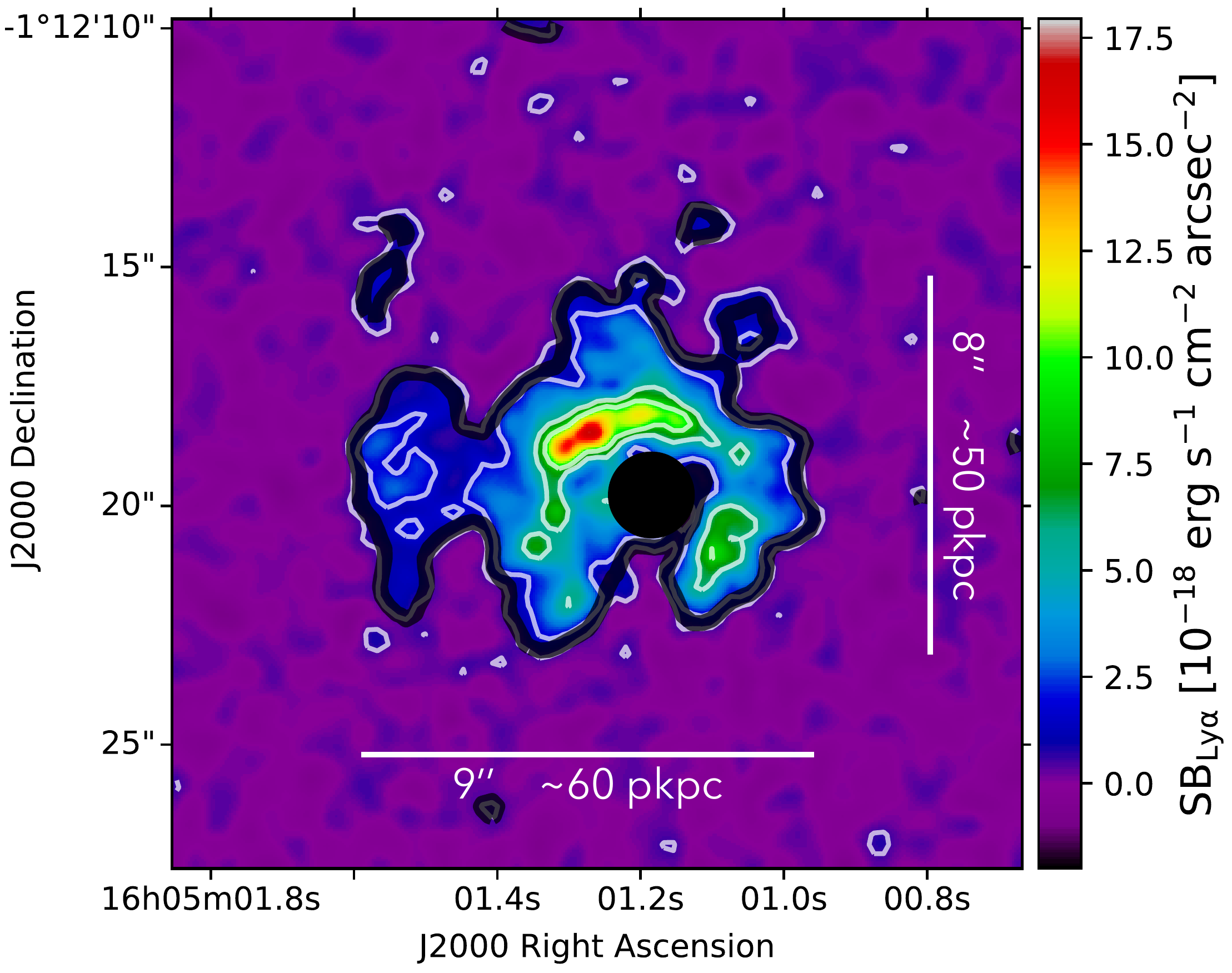}
	\caption{
		\textquotedblleft Optimally-extracted\textquotedblright~  image of the Ly$\alpha$ emission around J1605-0112. This image has been produced applying a 3D-mask (see discussion in Sec. \ref{sec:data_analysis}) to the PSF and continuum subtracted MUSE datacube. The position of the quasar is marked by a black dot. For display purposes, we have added - by means of the union operator - to the object 3D-mask one wavelength layer of the cube corresponding to the central wavelength of the nebulae.
		The spatial projection of the 3D-mask is indicated by the black thick contour that correspond to a SB of  $\rm \sim5\times10^{-19}erg~s^{-1} cm^{-2} arcsec^{-2}$. The thin contours indicate the propagated SNR in the images, with contour levels representing SNR$=2, 5, 10, 15$.
		The white horizontal (vertical) bar indicates a physical scale of $\sim$60 pkpc ($\sim$50 pkpc).
		Both horizontal and vertical bars are representative of the maximum linear projected size of the nebula on the RA and DEC axis. }
	\label{fig:Lya_flux}
\end{figure}

\begin{figure}
	\centering
	\includegraphics[width=1\columnwidth]{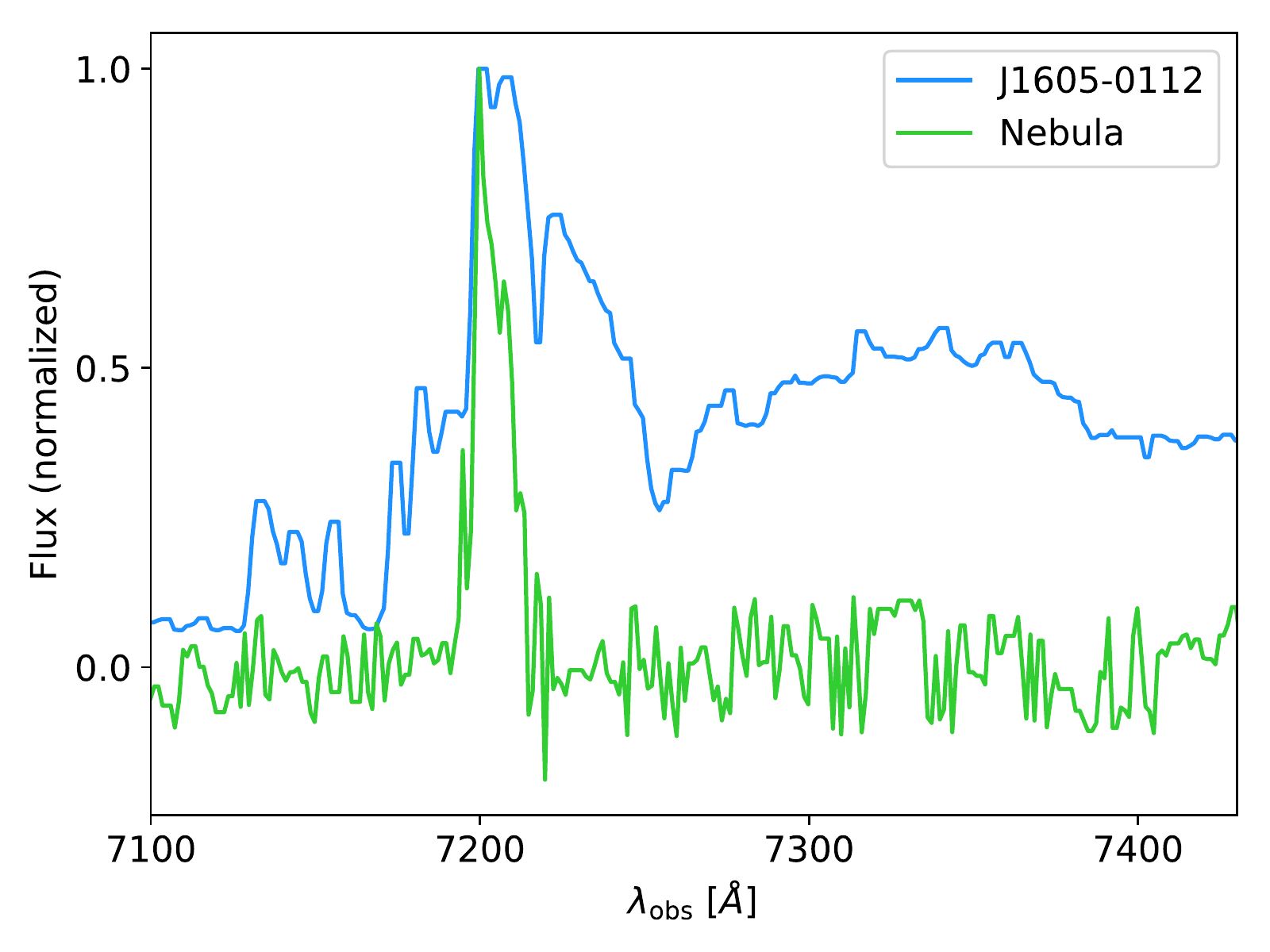}
	\caption{
		Comparison of the Ly$\alpha$ line shapes between the BAL quasar J1605-0112 (blue line) and the peak of the detected Ly$\alpha$ nebula (green line) surrounding the quasar.
		The spectrum of the nebula has been extracted within a circular aperture of $\sim 1''$ radius centered around the peak of its emission from the PSF and continuum subtracted final datacube.
		Both spectra have been normalized to their maximum to facilitate the comparison.
		The Ly$\alpha$ emission of the nebula appears to be much narrower than the broad  Ly$\alpha$ of the quasar J1605-0112, ensuring that the peak of the nebula close to the quasar position is not due to PSF residuals.
	}
	\label{fig:spectrum_comparison}
\end{figure}

Following the data processing steps described in Section \ref{sec:data_reduction} we detect a Ly$\alpha$ nebula, extended on circumgalactic scales around the quasar J1605-0112, at $z\sim4.92$. 
\newline
The nebula has been detected with a threshold of 10.000	connected voxels,  and the resulting \textquotedblleft optimally extracted image\textquotedblright, obtained using the 3D-segmentation-mask produced by \texttt{CubExtractor}, is shown in Fig. \ref{fig:Lya_flux}.
The black thick contour indicates the projection on the sky-plane of the 3D-mask, corresponding to a SB of $\rm \sim6\times10^{-19}erg~s^{-1} cm^{-2} arcsec^{-2}$.
A single spectral layer, corresponding to the central wavelength of the Ly$\alpha$, has been added - by means of the union operator - to the 3D-mask for display purposes.
\newline
The maximum spectral width defined in the 3D-mask corresponds to $\Delta\lambda = 38.75$ \AA. 
The total Ly$\alpha$ flux, obtained integrating the 3D-mask along the wavelength and the spatial directions, is $F_{\rm Ly\alpha} \sim \rm 1.7\times 10^{-16} erg~s^{-1} cm^{2}$, corresponding to a luminosity $L_{\rm Ly\alpha} \sim \rm 4.4\times 10^{43} erg~s^{-1}$. 
\newline
The Ly$\alpha$ emission projected on the sky-plane appears to extend over about 9 arcsec in RA and 8 arcsec in DEC, i.e., respectively 60 pkpc and 50 pkpc at this redshift.
The position of the quasar is marked by the black dot, whose dimension indicates the region used to normalize the PSF for each layer, i.e., the spatial region where the flux is dominated by the quasar (see Paragraph \ref{sec:psf_cont}).
This image has been obtained selecting all the voxels detected by the 3D-segmentation-mask  and integrating their fluxes along the wavelength direction.
\newline
Operationally speaking, this approach is equivalent to obtain various pseudo-NB images where the filter spectral size is optimized for each spaxel, to maximize the SNR of the object.
However, noise estimation for an image obtained trough this method is critical, as the noise associated to each spaxel depends on the number of layers that define it (i.e., the spectral size of pseudo-NB filter).
This results in the optimally-extracted flux map being affected by a noise that depends on the spatial position. 
To address this issue, following \cite{Borisova2016}, we estimated the noise and the SNR for each pixel in the map by propagating the variance and taking into account the number of layers contributing to each pixel.
Using propagated variances and the integrated flux we then created the SNR contours represented by the white thin contours in Fig. \ref{fig:Lya_flux}.
\newline
In Fig. \ref{fig:spectrum_comparison} we compare the Ly$\alpha$ profile of J1605-0112 with the Ly$\alpha$ profile of a region around the peak of the emission of the nebula (within an aperture of $\sim1''$).
The Ly$\alpha$ emitted by the nebula show a much narrower profile (FWHM $\sim 1000$ km s$^{-1}$) than the broad and complex  Ly$\alpha$ profile coming from the BAL quasar. 
Given such a difference between the widths and shapes of the spectra, we are confident that the Ly$\alpha$ emission from the nebula emission in close proximity to the quasar cannot constitute a PSF subtraction residual.
\newline  
\newline
It is worth stressing here that the \textquotedblleft optimally extracted image\textquotedblright~ of the Ly$\alpha$ emission shown in Fig. \ref{fig:Lya_flux} is different from a common standard NB image, because of the different number of spectral layers composing the pixels within the map.
This type of approach, enabled by the IFU technique, allows us to reveal both kinematically narrow and broad features that would have been either lost in the noise or underestimated in a NB image with a single, fixed width.
\newline
We also produced a \textquoteleft plain\textquoteright~ pseudo-NB image (i.e. by averaging a fixed number of spectral channels, without any tailoring) of the Ly$\alpha$ emission.
To this aim, we average over 38.75 \AA~ around the $\lambda_{\rm Ly\alpha}$ (rest-frame), that is the maximum spectral width of the nebula as defined by the 3D-mask (see Table \ref{tab:properties_nebula}).
Such a pseudo-NB image recovers $\sim80\%$ of the total flux revealed by the 3D-extraction, resulting in an apparent slightly smaller projected linear size of the nebula (about 6.5 arcsec in RA and 7 arcsec in DEC, respectively 42 pkpc and 45 pkpc at this redshift; see Appendix \ref{sec:appendix_nb}). 
For a consistent comparison with previous works, and to accurately characterize the noise and SB limits, we used the pseudo-NB image instead of the 3D optimally-extracted image to extract the circularly averaged surface brightness profile of the Ly$\alpha$ nebula, described in the next paragraph.
\begin{figure}
	\centering
	\includegraphics[width=1\columnwidth]{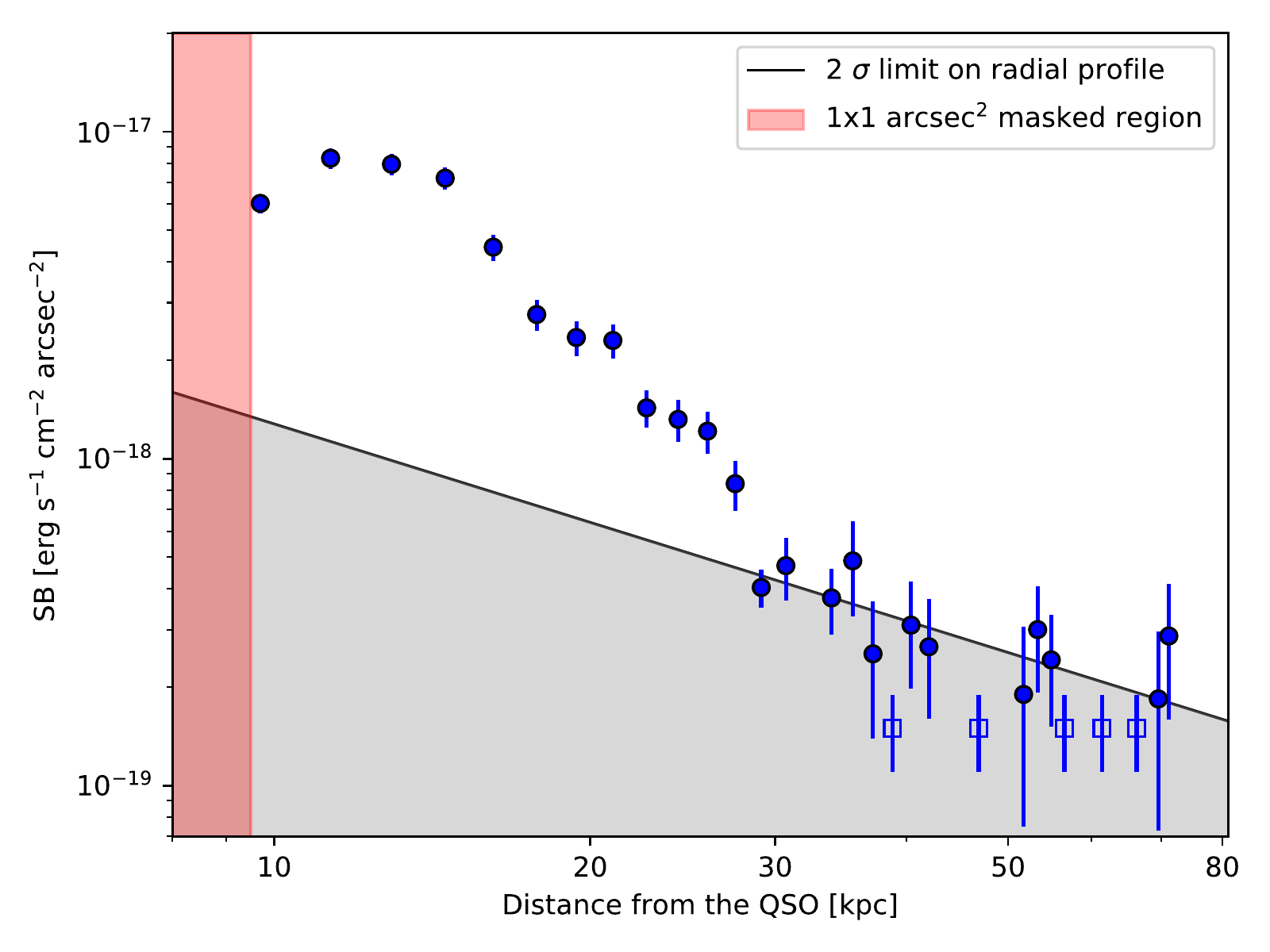}
	\caption{
		The circularly averaged SB profile (blue filled dots) of the Ly$\alpha$ nebula is shown as a function of the projected physical distance from J1605-0112. Open blue squares indicate negative values.
		Error bars represent 1$\sigma$ errors on the mean SB level measured in a given annulus.
		The SB profile has been extracted using a pseudo-NB image centered at the wavelength corresponding to the Ly$\alpha$ of the quasar, with a fixed width of 38.75\AA~ (see discussion in Sec. \ref{sec:SB_profile}). 
		The background flux level of the pseudo-NB image is consistent with zero. 
		The grey shaded area is an estimate of the 2$\sigma$  gaussian noise associated with the SB profiles.
		The pink shaded area is indicative of the 1x1 arcsec$^2$ region used to rescale the empirical PSF models (see Sec. \ref{sec:psf_cont}).
		}
	\label{fig:radial_profile_1}
\end{figure}
\begin{figure*}
	\centering
	\includegraphics[width=1\textwidth]{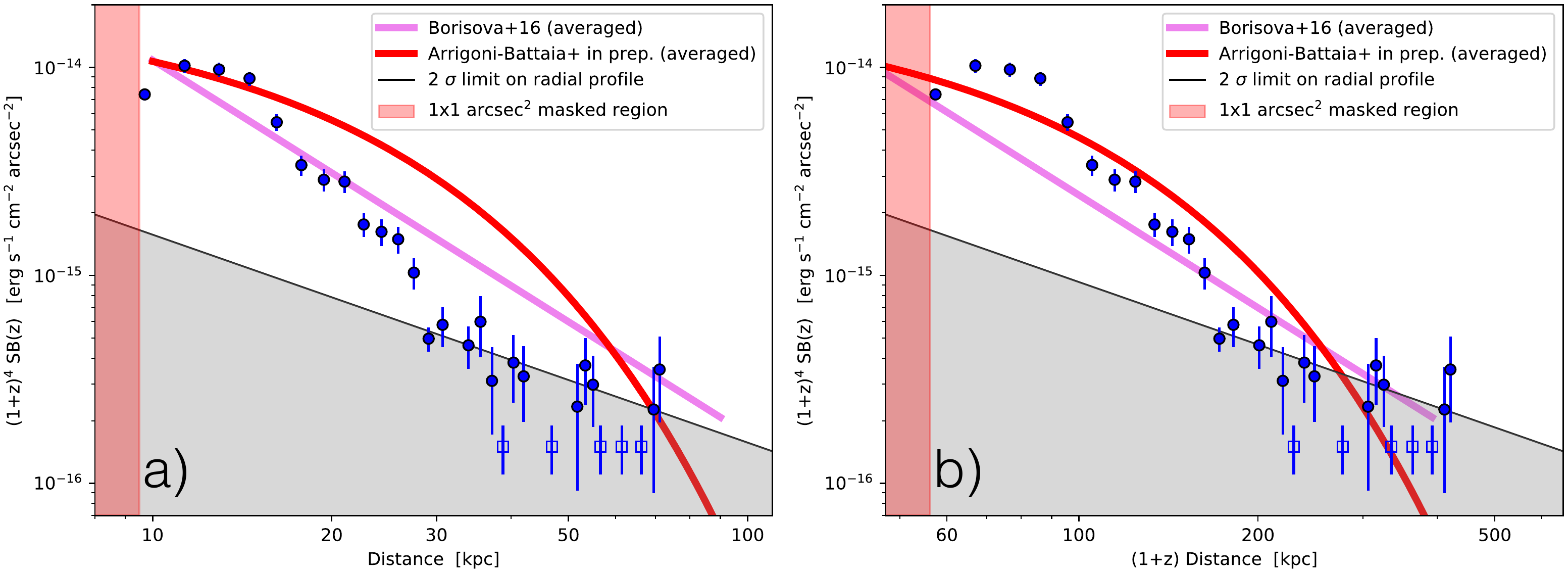}
	\caption{
		\textbf{a)}  The radial profile shown in Fig. \ref{fig:radial_profile_1} here is expressed in units of SB $\times (1+z)^4$ to avoid the cosmological SB dimming effect. Redshift-corrected averaged profiles of two samples of quasars observed with MUSE at mean redshift $z\sim3.2$ (\citealp{Borisova2016}; Arrigoni-Battaia et al., in prep.) are also shown for comparison. With a maximum linear projected size of $\sim30$pkpc from the quasar, our nebula appears to be intrinsically smaller than \textit{typical} nebulae observed around similar quasars at lower redshift. 
		\textbf{b)} The same radial profile is shown as a function of the projected comoving distance from the quasar, to take into account the cosmological growth of dark matter haloes. In this case, the intrinsic size discrepancy revealed in Fig. \ref{fig:radial_profile_2}a. is reduced (see Sec. \ref{sec:SB_profile} and Sec. \ref{sec:discussion} for an interpretation of such findings).
	}
	\label{fig:radial_profile_2}
\end{figure*}
\begin{figure*}
	\centering
	\includegraphics[width=1\textwidth]{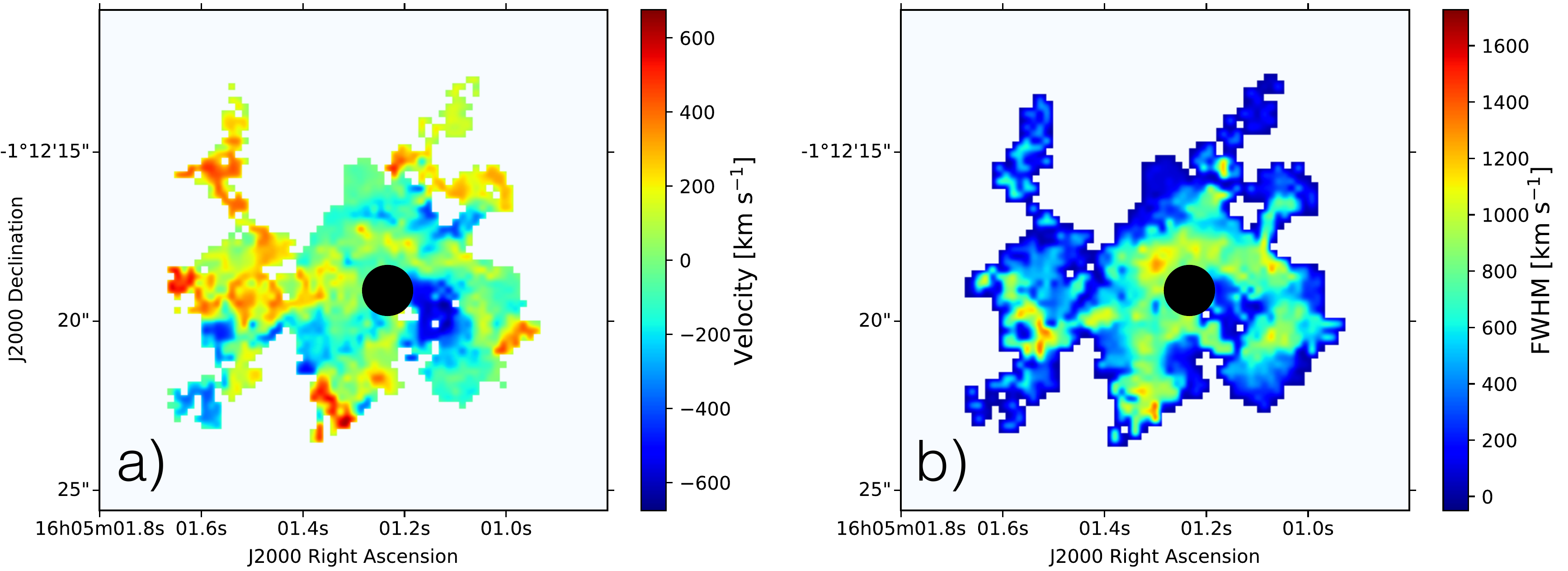}
	\caption{
		\textbf{a)} Velocity map traced by the Ly$\alpha$ emission around J1605-0112. The black circle indicates the position of the quasar. The velocity map does not show any clear evidence of rotation nor other coherent kinematic patterns, in line with similar recent MUSE observations at lower redshift (e.g., \citealp{Borisova2016}).
		\textbf{b)} Velocity dispersion map showed in units of Gaussian-equivalent FWHM (i.e., 2.35 times the velocity dispersion) for consistency with previous works in the literature. The black circle indicates the position of the quasar. The Ly$\alpha$ velocity dispersion map shows a particularly high broadening of the line (FWHM>1000 km s$^{-1}$), especially in the inner regions of the CGM, at $\gtrsim$10 pkpc from the quasar.
		}
	\label{fig:Lya_vel}
\end{figure*}

\subsection{Surface brightness profile}\label{sec:SB_profile}

We computed the circularly averaged SB profile of the Ly$\alpha$ nebula surrounding J1605-0112, at $z=4.92$, by using the pseudo-NB image of Ly$\alpha$, extracted as discussed above.
The resulting profile, as a function of projected physical distance from the quasar, is presented in Fig. \ref{fig:radial_profile_1}a (blue dots), where the expected 2-$\sigma$ gaussian SB limit for each aperture is also shown (grey shaded area).
To avoid the cosmological SB dimming effect and facilitate comparisons with similar objects at different redshifts, we computed the circularly averaged SB profile in units of SB$\times (1+z)^4$, i.e., cosmologically invariant brightnesses.
The resulting profile is shown in Fig. \ref{fig:radial_profile_2}a, where the redshift of the quasar, $z=4.92$, has been used as scaling value. 
Along with our results we also show the averaged profiles of two samples of quasars at redshift $z\sim3-4$ (the mean redshift is $z\sim3.2$ in both samples) by \cite{Borisova2016} and Arrigoni-Battaia et al. (in preparation) (violet and red solid lines respectively). 
Also these profiles are reported in cosmologically invariant SB units, using the mean redshift of the samples as scaling value. 
\newline
\newline
Both Fig. \ref{fig:radial_profile_1} and Fig. \ref{fig:radial_profile_2}a show that the circularly averaged SB radial profile of the nebula surrounding J1605-0112 extends up to $\sim30$ pkpc from the quasar; the mean flux measured in annuli at larger radii is consistent with the noise (2-$\sigma$) limit.
However,  Fig. \ref{fig:radial_profile_2}a shows that at distances consistent with the maximum extension of the radial profile of our nebula ($\sim30$ pkpc), the sensitivity provided by our deep MUSE observations would be enough to detect with high SNR ($\gtrsim5$) any  more extended emission, such as that expected by the averaged radial profiles of similar quasars at lower redshift.
\newline
This suggests that the Ly$\alpha$ nebula detected in the surrounding of J1605-0112, at $z\sim5$ is intrinsically smaller than typical Ly$\alpha$ nebulae observed with MUSE around luminous quasars at lower redshifts, $z\sim3-4$. 
We stress that such size discrepancy is intrinsic and not an artefact of the cosmological SB dimming; indeed the radial profiles have been expressed in invariant units, rescaling the quantities with redshift as discussed above. 
\newline
In Fig. \ref{fig:radial_profile_2}b we report the same SB radial profiles of Fig. \ref{fig:radial_profile_2}a, but in this case we show them as a function of the projected comoving distance from the quasar.
Using comoving distances, instead of physical, we can account for the cosmological growth of DM haloes, whose virial radii, for a fixed halo mass, scale as $\propto (1+z)^{-1}$ (\citealp{BarkanaLoeb2001}).
Fig. \ref{fig:radial_profile_2}b shows that the intrinsic size discrepancy between our nebula at $z\sim5$ and typical nebulae at $z\sim3-4$ is greatly reduced, or even disappears, when referring to comoving distances from the quasar. 
This suggests that our nebula appears to be less extended because haloes hosting quasars at high redshift
are, on average, intrinsically smaller.
In Sec. \ref{sec:discussion} we analyse possible origins and consequences of such finding.

\subsection{Kinematics}\label{sec:Kinematics}

We used the 3D-segmentation-mask, defined in Sec. \ref{sec:data_analysis}, to also produce two-dimensional maps of the first and second moments of the flux distribution in its spectral domain.
Such maps give indication of, respectively, the centroid velocity and the width of the emission line (the velocity dispersion) at each spatial location.
\newline
In Fig. \ref{fig:Lya_vel}a we report the map of the first moment of the Ly$\alpha$ flux distribution, i.e., the shift of the velocity centroid with respect to the peak of the integrated emission of the nebula.                                                                                                        
The velocity map of Ly$\alpha$ emission does not show any clear evidence of rotation nor any coherent kinematic patterns, in line with similar recent MUSE observations at lower redshift (\citealp{Borisova2016}).
However we note that, due to its resonant nature, Ly$\alpha$  is usually not a good tracer of kinematics as any coherence on 10 pkpc scales may be disrupted by radiative transfer effects (\citealp{Cantalupo2005}).
\newline
In Fig. \ref{fig:Lya_vel}b we present the map of the second moment of the Ly$\alpha$ flux distribution, i.e., the velocity dispersion (for consistency with previous works in the literature we show the Gaussian-equivalent FWHM derived by multiplying the second moment by 2.35). 
The Ly$\alpha$ velocity dispersion map shows a particularly high broadening of the line (FWHM>1000 km s$^{-1}$), especially in the inner regions of the CGM, at $\sim$10 pkpc from the quasar.
Such a high velocity dispersions are in contrast with previous MUSE observations by \cite{Borisova2016}, which found narrower nebulae with FWHM$\sim$500-700 km s$^{-1}$  surrounding 16 (out of 17) radio-quiet quasars, and are more similar to what found in nebulae around radio-loud systems (\citealp{Borisova2016}).
Two notable exceptions are the nebula \#6, surrounding J0124+0044 at $z\sim3.8$ (\citealp{Borisova2016}), and the Enormous Lyman-Alpha Nebula (ELAN) surrounding  J1020+1040 at $z=3.164$ (\citealp{ArrigoniBattaia2017}), both showing a broad Ly$\alpha$ emission (FWHM $\sim 1000$ km s$^{-1}$) on similar scales ($\sim 10$ kpc from the quasar) as for our nebula.

\subsection{CIV/Ly$\alpha$ line ratio}\label{sec:line_ratio}
After hydrogen Ly$\alpha$ line, CIV $\lambda$1549 \AA~and HeII $\lambda$1640 \AA~are the brightest UV lines found within quasar-powered nebulae, and their line ratios have been often used to constrain the ionisation parameter, gas density and metallicity, providing information on the origin and physical properties of the emitting gas (see \citealp{Nagao2006, ArrigoniBattaia2015, ArrigoniBattaia2015b, Prescott2015}, and references therein).
For instance, giant Ly$\alpha$ nebulae around high-$z$ radio galaxies typically show high CIV/Ly$\alpha$ and HeII/Ly$\alpha$ (ratios $\gtrsim0.1$; see \citealp{VillarMartin2006}), which are currently explained by high metallicity gas whose emission is driven by ionization and jet-ambient interaction.
We searched only for possibly extended CIV within our MUSE datacube. 
We note that, at $z=4.92$, the MUSE wavelength coverage does not enable us to study the HeII emission line. 
In order to search for spatially extended CIV emission we used the Ly$\alpha$-defined 3D-mask produced by \texttt{CubExtractor}. Following the analysis of \cite{Borisova2016} we scanned spectral layers around the location of CIV, by shifting the mask along the wavelength direction of the PSF and continuum subtracted datacube.
Using the same 3D-mask defined for the Ly$\alpha$ avoids any aperture effects in the line ratios. 
As the Ly$\alpha$ is expected to be intrinsically brighter and more extended than CIV, this approach gives us very conservative limits in case of non-detection.
Considering the possibility of radiative transfer-induced velocity shift between the Ly$\alpha$ line emission and the CIV we have scanned spectral layers within a window of 3000 km s$^{-1}$ on both side of the expected CIV line. 
We then compute the total flux of the voxels associated with the 3D-mask at each spectral location.
\newline
With this approach we do not detect any CIV extended emission around J1605-0112, within a 1$\sigma$ limit of  $\rm SB_{CIV}$ (1$\sigma$) $\rm \sim4.5\times10^{-18}~erg~s^{-1} ~cm^{-2}$.
This yields a 2$\sigma$ upper limit on the CIV/Ly$\alpha$ ratio of $\sim$0.06, consistently with what found in the majority of radio-quiet quasars at lower redshifts (e.g., \citealp{Borisova2016}).
In the Appendix \ref{sec:appendix_CIV}, we show a 30\AA-width NB image of the CIV, clearly showing that the CIV emission is undetected.

\section{Discussion}\label{sec:discussion}

\subsection{The nebula small size relative to lower redshift nebulae observed with MUSE}

With a maximum projected linear size of $\sim$60 physical kpc (pkpc), the Ly$\alpha$ nebula around J1605-0112, at $z\sim5$, appears to be smaller than typical nebulae ubiquitously detected with MUSE around 
quasars at lower redshifts ($z\sim3-4$), extended on linear sizes larger than 100 pkpc, up to $\sim$300 pkpc (see \citealp{Borisova2016}).
\newline
As discussed in Sec. \ref{sec:SB_profile}, we verified that such a size discrepancy is not a trivial effect of the surface brightness (SB) cosmological dimming. 
Fig. \ref{fig:radial_profile_2}a shows that, after correcting for the cosmological dimming effect (i.e., by expressing SB in units of SB$\times(1+z)^4$), our $\sim5$h on-source of MUSE integration provides a sensitivity level such that we should detect with high SNR ($\gtrsim5$) any \textit{typical} emission at physical distances $\gtrsim30$ pkpc from the quasar, which would be expected by the averaged radial profiles of similar nebulae at lower redshift. 
In the following we provide two different scenarios whose combination is most likely the correct way to interpret such a physical size discrepancy.
\newline
\newline
Cosmological theories of structure formation and, more in general the $\Lambda$CDM paradigm, predict that dark matter (DM) haloes collapsing at redshift $z$ have a (physical) virial radius 
$r_{vir} \propto M^{1/3} (1+z)^{-1}$ (e.g., \citealp{BarkanaLoeb2001}).
Thus, DM haloes at higher redshifts, if they remain in the same mass range, are smaller than DM haloes at lower redshifts.
In Fig. \ref{fig:radial_profile_2}b we showed the same SB dimming-corrected radial profiles shown in  Fig. \ref{fig:radial_profile_2}a, but as a  function of projected comoving distance from the quasars.
Expressing distances in comoving units, defined as $d_{comov} = d_{phys} \times (1+z)$, compensates the redshift evolution of virial radii accounting for the cosmological growth of DM haloes across the cosmic time.
Fig. \ref{fig:radial_profile_2}b shows that, when referring to comoving distances, the intrinsic size discrepancy between our nebula at $z\sim5$ and typical nebulae at $z\sim3-4$ is greatly reduced.
This suggests that our nebula appears to be less extended because quasar hosting DM haloes at high redshift are smaller than haloes at lower redshifts, assumed that they are hosted on average by haloes with comparable masses in the redshift range $3-5$.
\begin{figure}
	\centering
	\includegraphics[width=1\columnwidth]{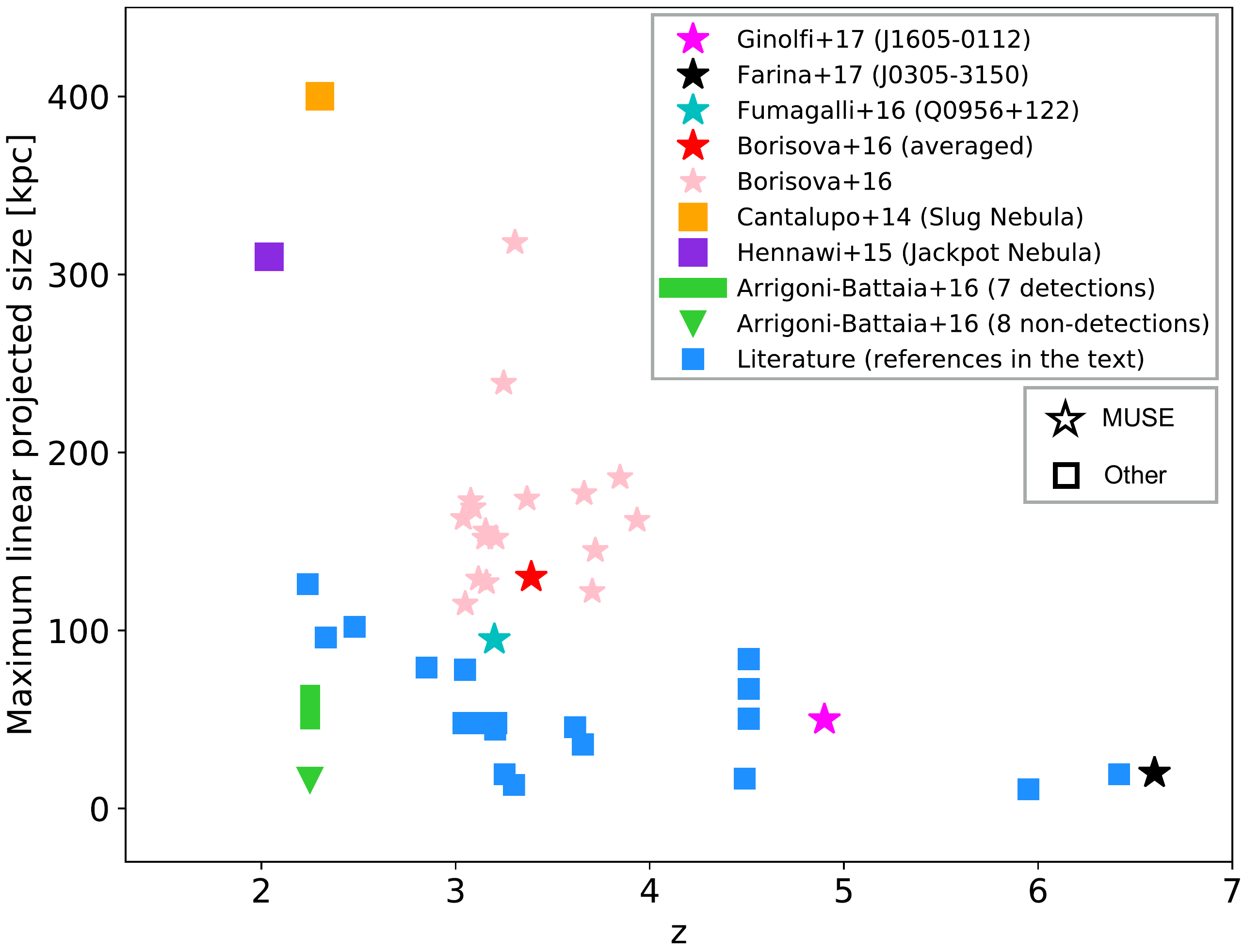}
	\caption{The maximum projected sizes of a compilation of Ly$\alpha$ Nebulae surrounding radio-quiet quasars is shown as a function of redshift. 
	Star symbols refer to observations taken with MUSE, while squares to observations taken with narrow band filter technique (triangles refer to non-detections).
	Although the sample is affected by some caveats (see Sec. \ref{sec:discussion}), this qualitative analysis suggest a positive trend between the sizes of Ly$\alpha$ nebulae and the cosmic time within the redshift range $z\sim2-7$.
	}
	\label{fig:trend}
\end{figure}
\newline
\newline
Such a result would suggest a physical correlation between the size of Ly$\alpha$ nebulae and the sizes of DM haloes/structures around quasars. 
This conjecture is supported by the qualitative analysis of other observations of Ly$\alpha$ nebulae in an even wider redshift range. \newline
In Fig. \ref{fig:trend} we report the sizes (maximum linear projected sizes) of a compilation%
\footnote{We updated the collection reported in the PhD thesis of Arrigoni-Battaia  (Table 1.2; \url{https://www.imprs-hd.mpg.de/49473/thesis_Arrigoni.pdf}). 
The updated list includes data from:	
\cite{Heckman1991b, Heckman1991a, Bremer1992, Roettgering1997,  vanOjik1997, Lehnert1998,Bergeron1999, Fynbo2000, Bunker2003, Weidinger2004, Weidinger2005, Christensen2006, Barrio2008, Courbin2008,Smith2009,Goto2009, Matsuda2011, Willott2011, Adelberger2005, North2012, Goto2012, Humphrey2013, Roche2014, Cantalupo2014, Hennawi2015, ArrigoniBattaia2016, Borisova2016, Fumagalli2016, Farina2017}.} 
of Ly$\alpha$ nebulae (including non-detections) surrounding radio-quiet quasars as a function of redshift, suggesting a positive trend between the sizes of Ly$\alpha$ nebulae and the cosmic time within the explored redshift range, i.e., $z\sim2-7$.
This qualitative trend has to be taken with caution due to the following caveats:
(i) maximum projected sizes have been measured through different methods and with different SB limits, (ii) many of the nebulae are asymmetrical and it is not obvious how to characterize them with a single physical size, (iii) the sample is biased towards detections, as many non-detections have not been published.
We leave a more quantitative analysis of this trend, including a re-processing of data in the literature and a self-consistent calculation of the maximum projected sizes (using a common isophote reference) for a future work.
Fig. \ref{fig:trend} also shows an interesting increase in the scatter across the cosmic time, possibly due to a lack of observations for high-redshift quasars or, more interestingly, to galaxy formation processes: quasar host haloes at low-$z$ have a larger variety of properties (e.g., masses, environment, amount of cold gas, UV source luminosities) than their higher-$z$ counterparts.
\newline
\newline
Another interpretation, expected to contribute to the observed size discrepancy, may be ascribed to the BAL-nature of the quasar J1605-0112.
BAL quasars are characterized by redder continua with respect to non-BAL quasars (see \citealp{Maiolino2004}). 
In particular J1605-0112 requires a reddening $\rm E_{B-V} = 0.03$ of SMC extinction curve in order to match the observed near-IR continuum slope (\citealp{Reichard2003, Maiolino2004}).
This high reddening is often related to a high column density of dust in the outflowing circumnuclear medium around the Active Galactic Nucleus (AGN), able to efficiently absorb the AGN-powered UV photons, able to suppress the escape fraction of the ionizing radiation necessary to induce fluorescence in the neutral circumgalactic gas. 

\subsection{The origin of the high  Ly$\alpha$  broadening}

Although Ly$\alpha$, due to its resonant nature, is not generally considered a good tracer of kinematics, the analysis of circumgalactic gas kinematics of the nebula surrounding J1605-0112, reported in Sec. \ref{sec:Kinematics}, shows peculiar results. 
In particular the velocity dispersion map reported in Fig. \ref{fig:Lya_vel}b (where the gaussian-equivalent FWHM of the line is shown at each spatial location) shows a  particularly high broadening of the line (FWHM > 1000 km s$^{-1}$), especially in the inner regions of the CGM, at $\sim$10 kpc from the quasar, in agreement with the scales in \cite{ArrigoniBattaia2017}.
Such a high value of FWHM exceeds by a factor of $\sim2$ the typical FWHM observed in other works.
For instance \cite{Borisova2016} reported observations of narrower nebulae, e.g., with FWHM$\sim$500-700 km s$^{-1}$, surrounding almost all the radio-quiet quasars in their sample (with the notable exception of the nebula \#6, surrounding J0124+0044 at $z\sim3.8$, showing a radial profile and a kinematics that looks similar to our nebula).
\newline
A possible interpretation of such high broadening of the Ly$\alpha$ line emission in the inner region of our nebula may be ascribed to the BAL nature of the quasar J1605-0112 (\citealp{Maiolino2004}). 
Indeed, the outflow inferred from the BAL features, may not be restricted to the circumnuclear region, but extend to large scales. 
For instance, based on a detailed spectral analysis of a few BAL quasars, \cite{Borguet2012} infer that in some cases the outflowing gas traced by the absorption troughs is distributed on scales of 10-30 kpc, hence revising the standard scenario in which BALs are only associated with outflows on small scales. 
Therefore the large FWHM of the Ly$\alpha$ may directly trace the large scale outflow associated with the BAL. 
Even if the Ly$\alpha$ is not directly tracing the outflow gas, as suggested by the null detection of the CIV line, possibly pointing to a low gas metallicity, the outflow can anyhow introduce significant turbulence into the CGM which may result in the observed broadening.

\section{Conclusions}\label{sec:conclusions}
Recent MUSE observations have revealed ubiquitous  giant Ly$\alpha$ nebulae around bright quasars at $z\sim3-4$, extending up to $\sim300$ pkpc (e.g., \citealp{Borisova2016, Fumagalli2016, ArrigoniBattaia2017}).
In this work we present deep ($\sim 4$ h) MUSE observations of J1605-0112, a Broad Absorption Line (BAL) quasar at higher redshift, $z\sim5$. 
Our findings can be summarized as follows:
\newline
\newline
(i) we reveal a Ly$\alpha$ nebula around J1605-0112. The Ly$\alpha$ emission projected on the sky-plane appears to extend over $\sim8-9$ arcsec, i.e., $50-60$ kpc at $z=4.92$ (Fig. \ref{fig:Lya_flux});
\newline
\newline
(ii) after correcting for the cosmological surface brightness (SB) dimming, we compared the circularly averaged SB radial profile of our nebula with the averaged radial profiles of other nebulae at lower redshift, observed with MUSE. 
The profile of our Ly$\alpha$ nebula, at $z\sim5$, shows a steeper slope than typical nebulae at lower redshifts and indicate a less extended gas distribution (Fig. \ref{fig:radial_profile_2}a).
It is important to note, however, that single nebulae at lower redshift show a large variability, also in terms of radial profiles;
\newline
\newline
(iii) the effect described in (ii) is greatly reduced when referring to comoving distances (Fig. \ref{fig:radial_profile_2}b), which take into account the cosmological growth of dark matter (DM) haloes (whose virial radius is expected to evolve proportionally to $1/(1+z)$). 
This suggests that the discrepancy between the Ly$\alpha$ extensions may be ascribed
to the smaller size of DM haloes with comparable mass at higher $z$, implying an interesting empirical relation between the size of Ly$\alpha$ nebulae and sizes of DM haloes around quasars.
This conjecture is supported by a qualitative analysis of other observations of Ly$\alpha$ nebulae in a larger redshift range (see Fig. \ref{fig:trend});
\newline
\newline
(iv) while the velocity map does not show any clear pattern (Fig. \ref{fig:Lya_vel}a; in line with similar observations at lower redshift), the velocity dispersion map shows a particularly high broadening of the line (FWHM > 1000 km s$^{-1}$) in the inner regions of the nebula (Fig. \ref{fig:Lya_vel}b).
\newline
\newline
We suggest that (iii) and (iv) findings may be favoured by the BAL nature of the observed quasar.
Indeed BAL quasars are characterized by redder continua with respect to non-BALs (e.g., \citealp{Reichard2003, Maiolino2004}) and this is often associated with high dust column densities in the circumnuclear regions around the AGN. 
Such a large dust content may efficiently absorb the UV photons emitted by the AGN and suppress the escape fraction of the ionizing radiation responsible for producing the Ly$\alpha$ fluorescence in the CGM, thus reducing the detectable size of Ly$\alpha$ nebulae. \newline
Moreover, BALs have been observed to experience powerful outflows and in particular LoBALs (as J1605-0112) show extremely deep troughs of low-ionization lines (e.g., \citealp{Maiolino2004}), indicative of very large columns of outflowing gas. 
This is consistent with BALs tracing a phase of quasar feedback particularly effective in ejecting gas (e.g., \citealp{Dunn2010}). 
In this light, the broad Ly$\alpha$ emission observed in the inner 
region (on $\sim$10 kpc scales) around our BAL quasar may trace both highly turbulent circumgalactic gas as well as outflowing material escaped from the galaxy. 

\section*{Acknowledgements}
The authors would like to thank the anonymous referee for the useful comments.
M. Ginolfi thanks F. G. Saturni for helpful discussions.
The research leading to these results has received funding from the European Research Council (ERC) under the European Union's Seventh Framework Programme (FP/2007-2013) / ERC Grant Agreement n. 306476.
R. Maiolino acknowledges support from the ERC Advanced Grant 695671 \textquoteleft QUENCH\textquoteright. 
R. Maiolino and S. Carniani acknowledge support from the Science and Technology Facilities Council (STFC).
S. Cantalupo gratefully acknowledges support from Swiss National Science Foundation grant PP00P2\_163824.
The work is based on observations collected at the European Organisation for Astronomical Research in the Southern Hemisphere under ESO programme 095.B-0875(A).
This research made use of \texttt{ASTROPY}, a community-developed core \texttt{PYTHON} package for Astronomy (\citealp{Astropy2013}).

\bibliographystyle{mnras}
\bibliography{biblio} 


\appendix

\section{MUSE spectrum of J1605-0112 and its BAL features}\label{sec:MUSE_spectrum}
\begin{figure*}
	\centering
	\includegraphics[width=0.6\textwidth]{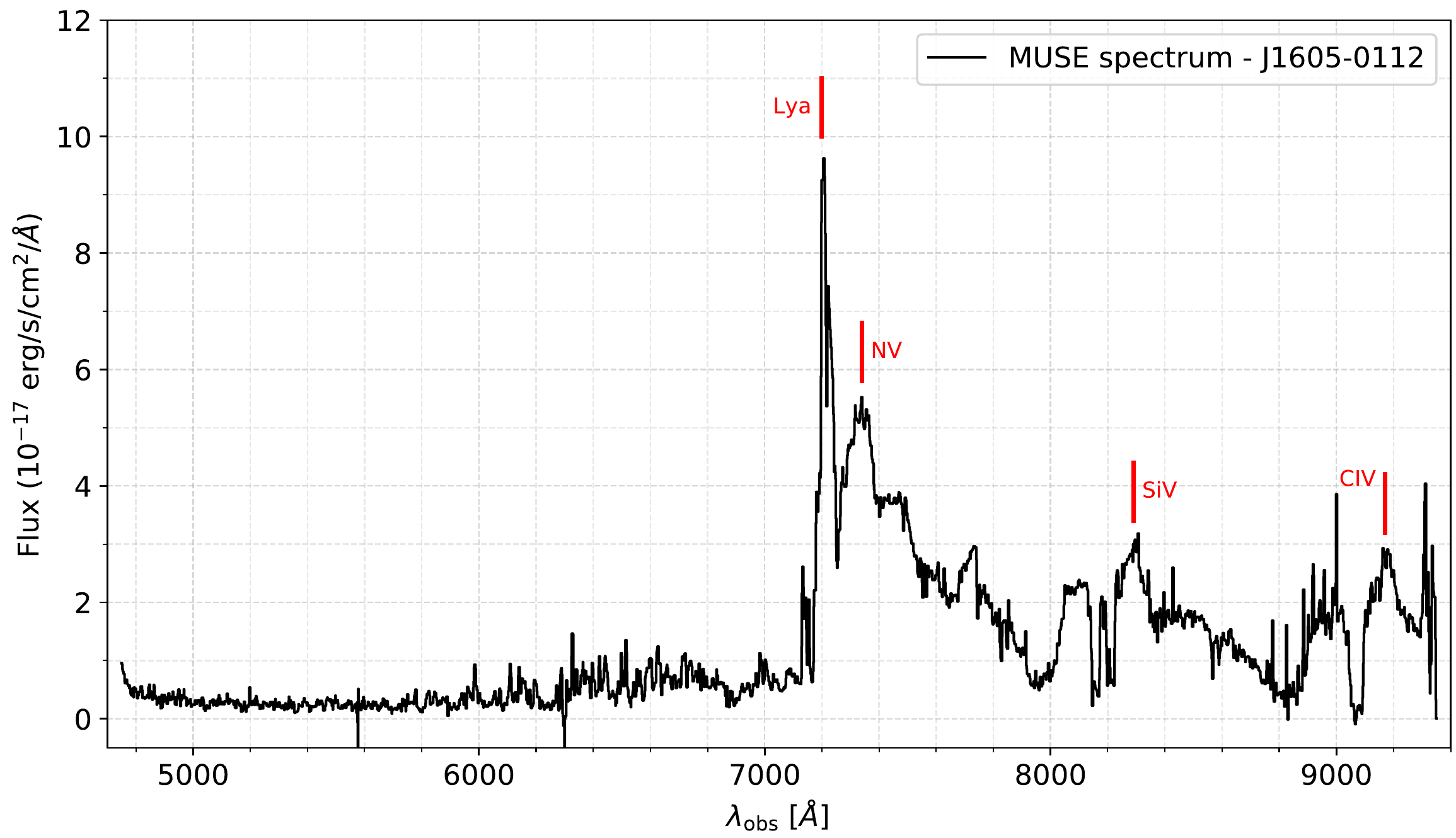}
	\caption{
		One-dimensional spectrum of J1605-0112 obtained with MUSE. 
		The spectrum has been extracted in a 1.5 arcsec radius circular aperture, containing most of the flux.
		The red lines indicate the location of UV emission lines, e.g., Ly$\alpha$, NV, SiV, CIV.}
	\label{fig:MUSE_spectrum}
\end{figure*}

In Fig. \ref{fig:MUSE_spectrum} we show the 1D spectrum of the BAL quasar J1605-0112, at $z=4.92$.
The spectrum has been extracted in a 1.5 arcsec radius circular aperture, containing most of the flux, consistently with SDSS.
\newline
As visible from the spectrum, the deepest trough of the CIV, at $\lambda_{\rm{obs}}\sim9070$ \AA,~absorbs almost completely the light emitted by the quasar.
In particular, using the spectral information of our MUSE observations, after computing a linear fit of the continuum emission in the wavelength range of the deep CIV trough, we estimate that the absorption leaves a residual flux which is only about $3\%$ of the continuum emission at those wavelengths.
Such a residual light in the deepest absorption trough is consistent with a BAL covering fraction of the order of unity (e.g., \citealp{Elvis2000}).
It is unlikely that such 3\% residual emission is associated with the host galaxy as the radial profile is consistent with what observed in the quasar continuum outside the trough.
Applying the same analysis to the other troughs of CIV, SiV and NV, we found less extreme residual fluxes, ranging from 20\% to 60\%.
\newline
A more detailed analysis of the emission in the BAL troughs is beyond the scope of this paper and we leave it  for future works.

\section{white-light image}\label{sec:white-light}
\begin{figure}
	\centering
	\includegraphics[width=0.7\columnwidth]{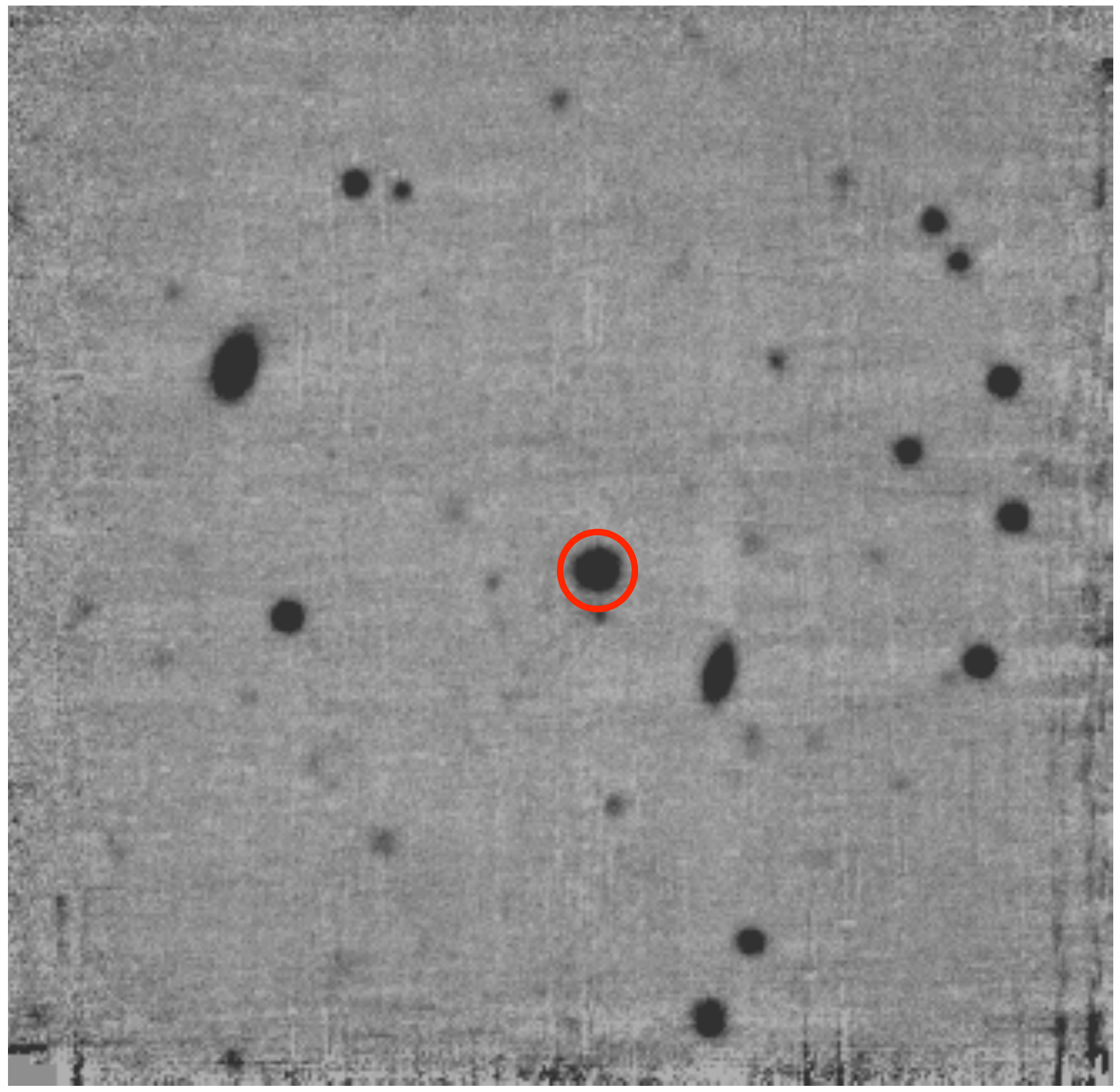}
	\caption{
		White-light image of the whole 1x1 arcmin$^2$ FOV of the final datacube.
		The red circle indicates the position of the quasar.}
	\label{fig:whitelight}
\end{figure}

In Fig.  \ref{fig:whitelight} we show the whole 1x1 arcmin$^2$ field of view of the final datacube as a \textquotedblleft white-light\textquotedblright~ image, obtained by collapsing the datacube along the wavelength direction.

\section{pseudo-NB image of the Ly$\alpha$ emission}\label{sec:appendix_nb}

\begin{figure}
	\centering
	\includegraphics[width=0.9\columnwidth]{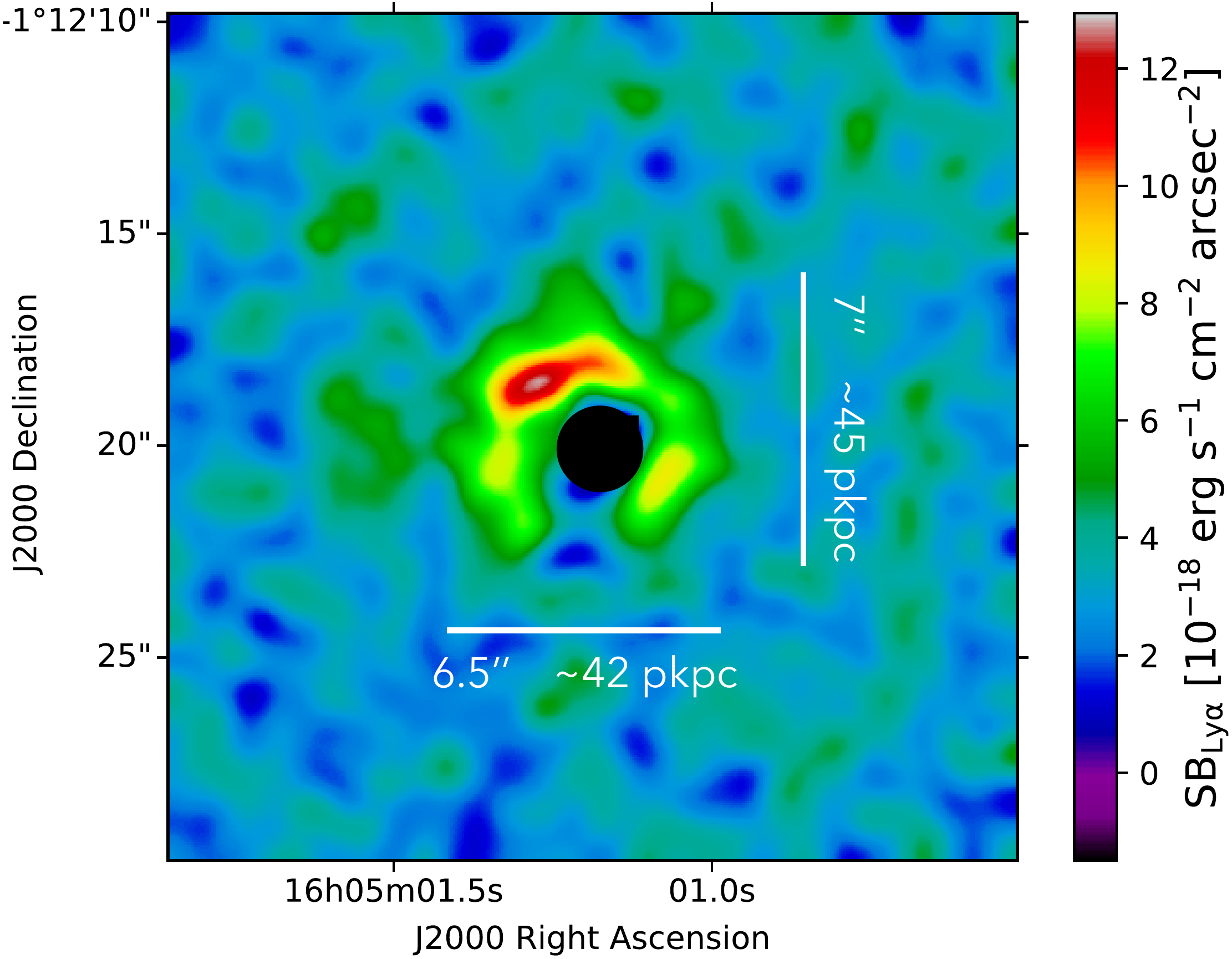}
	\caption{Pseudo-NB image of the Ly$\alpha$ emission around J1605-0112, centered at the wavelength corresponding to the Ly$\alpha$ of the quasar, with a width coincident with the maximum spectral width of the nebula as defined by the 3D-mask, i.e., 38.75\AA. 
	To facilitate the comparison, the data are shown on the same scale of Fig. \ref{fig:Lya_flux}.
	}
	\label{fig:NB}
\end{figure}

In Fig. \ref{fig:NB} we show a pseudo-NB image of the Ly$\alpha$ emission, obtained by averaging 38.75 \AA~ (i.e., the maximum spectral width of the nebula as defined by the 3D-mask) around the rest-frame wavelength of Ly$\alpha$ (see Table \ref{tab:properties_nebula}).

\section{Non-detection of CIV emission}\label{sec:appendix_CIV}
\begin{figure}
	\centering
	\includegraphics[width=1\columnwidth]{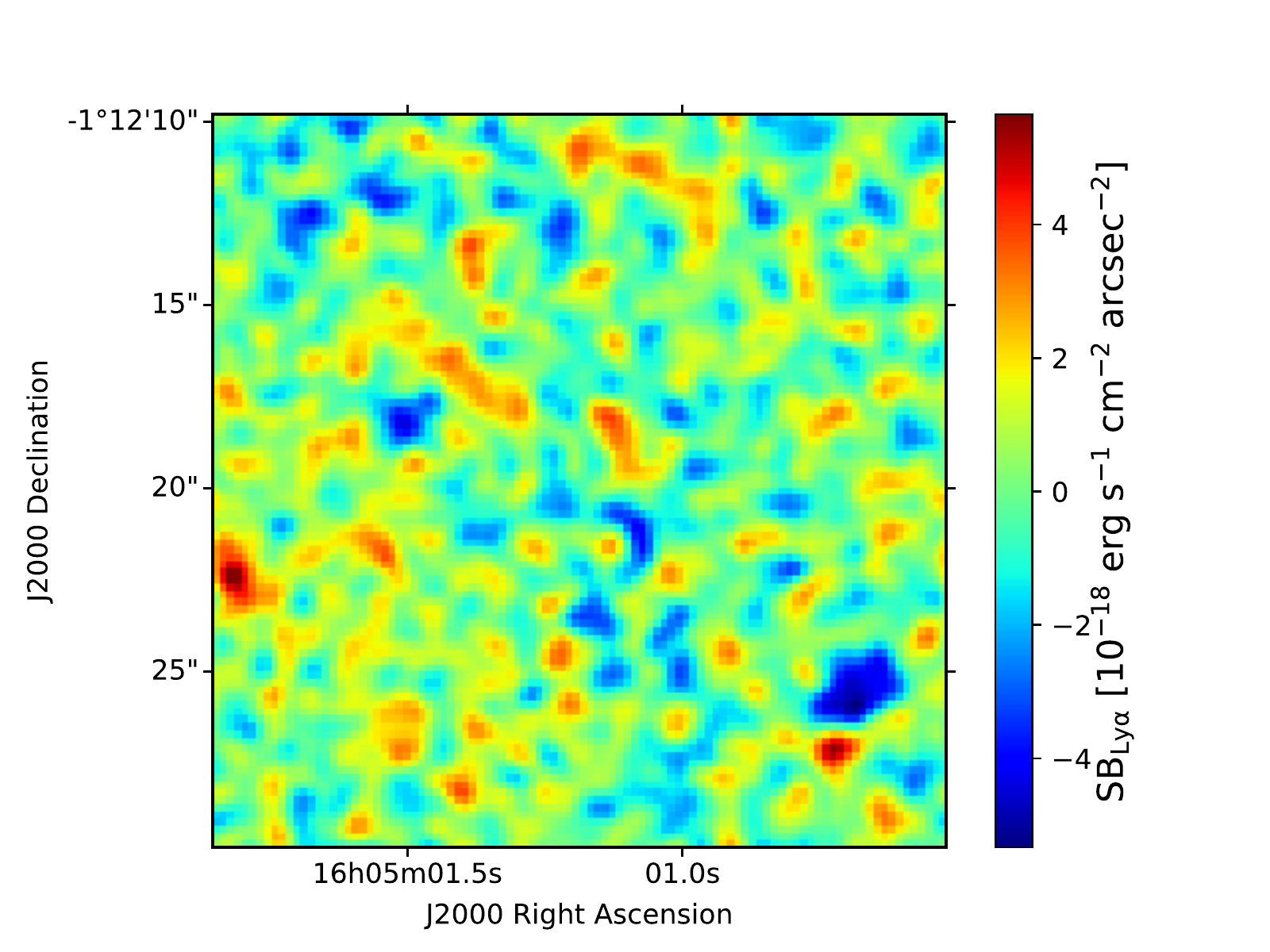}
	\caption{NB image of the CIV emission around J1605-0112, obtained by averaging 30\AA~around the CIV central wavelength (rest-frame 9170\AA). The CIV emission is clearly undetected.}
	\label{fig:CIV}
\end{figure}

In Fig. \ref{fig:CIV} we show a pseudo-NB image of the CIV emission, obtained by averaging 30\AA~around the CIV central wavelength (rest-frame 9170\AA).
It is evident that the CIV emission is undetected.

\bsp	
\label{lastpage}
\end{document}